\documentclass[prb,twocolumn,superscriptaddress,floatfix,showpacs]{revtex4}
\usepackage{graphicx,amsfonts,amssymb,amsmath,hyperref,enumerate}

\newif\ifhyper
\hypertrue
\ifhyper
\hypersetup{
   citecolor = {green},
   colorlinks = {true}, 
   urlcolor = {blue} 
}
\fi

\newcommand{\beq}{\begin{equation}}
\newcommand{\eeq}{\end{equation}}
\newcommand{\beqa}{\begin{eqnarray}}
\newcommand{\eeqa}{\end{eqnarray}}
\newcommand{\ket} [1] {\vert #1 \rangle}
\newcommand{\bra} [1] {\langle #1 \vert}

\def\bra#1{\langle#1\vert}
\def\ket#1{\vert#1\rangle}

\def\Longarrow{\protect\@lra}
\def\@lra{\relbar\joinrel\relbar\joinrel\relbar\joinrel%
          \relbar\joinrel\rightarrow}

\begin{document}

\title{Symmetry-protected intermediate trivial phases in quantum spin chains}

\author{Augustine Kshetrimayum}
\affiliation{Institute of Physics, Johannes Gutenberg University, 55099 Mainz, Germany}

\author{Hong-Hao Tu}
\affiliation{Max-Planck-Institut f\"ur Quantenoptik, Hans-Kopfermann-Str.~1, 85748 Garching, Germany}

\author{Rom\'an Or\'us}
\affiliation{Institute of Physics, Johannes Gutenberg University, 55099 Mainz, Germany}
\date{20th October 2015, updated}
\begin{abstract}
Symmetry-protected trivial (SPt) phases of matter are the product-state analogue of symmetry-protected topological (SPT) phases. This means, SPt phases can be adiabatically connected to a product state by some path that preserves the protecting symmetry. Moreover, SPt and SPT phases can be adiabatically connected to each other when interaction terms that break the symmetries protecting the SPT order are added in the Hamiltonian. It is also known that spin-1 SPT phases in quantum spin chains can emerge as effective intermediate phases of spin-2 Hamiltonians. In this paper we show that a similar scenario is also valid for SPt phases. More precisely, we show that for a given spin-2 quantum chain, effective intermediate spin-1 SPt phases emerge in some regions of the phase diagram, these also being adiabatically connected to non-trivial intermediate SPT phases. We characterize the phase diagram of our model by studying quantities such as the entanglement entropy,  symmetry-related order parameters, and 1-site fidelities. Our numerical analysis uses Matrix Product States (MPS) and the infinite Time-Evolving Block Decimation (iTEBD) method to approximate ground states of the system in the thermodynamic limit. Moreover, we provide a field theory description of the Êpossible Êquantum phase transitions between the SPt phases. Together with the numerical results, such a description shows that the transitions may be described by Conformal Field Theories (CFT) with central charge $c=1$. Our results are in agreement, and further generalize, those in [Y. Fuji, F. Pollmann, M. Oshikawa, Phys. Rev. Lett. 114, 177204 (2015)]. 

\end{abstract}

\pacs{75.10.Pq, 75.10.Jm}

\maketitle

\section{Introduction}
\label{sec1}

The study of quantum many-body entanglement has given us many lessons. One of them, and quite important, is the fact that gapped phases of matter can either be topologically ordered or trivial  in the absence of protecting symmetries,  depending on whether there is long-range entanglement in the ground state or not. Moreover, such phases may be protected or enriched by specific symmetries present in the many-body Hamiltonian. A prominent example is the well-known concept of symmetry-protected topological order (SPT) \cite{wen0}. ÊEven though such phases do not have intrinsic topological order, 
they cannot be distinguished by local order parameters, and hence fall beyond the paradigm of Landau's theory of phase transitions \cite{landau}. Focusing on one-dimensional (1d) systems, SPT phases have interesting properties such as hidden string order, fractionalized gapless edge modes, and degeneracy in the entanglement spectrum, just to name a few \cite{SPT}. Such properties are protected by the symmetries, i.e., they cannot be destroyed unless the symmetries are broken by some terms in the Hamiltonian. Partly because of this robustness, SPT phases have been proposed as a resource for quantum information processing, e.g., as quantum repeaters \cite{rep}, or as the substrate for measurement-based quantum computation \cite{qcom}. 

In 1d, the most famous example of an SPT phase is the spin-1 Haldane phase \cite{haldane1}, of which the AKLT state is the paradigmatic  representative \cite{AKLT}. As already explained extensively in the literature (see, e.g., Ref.\cite{SPT} and references therein), this gapped phase is protected by either one of time reversal ($T$), bond-centered inversion ($I_b$), or rotation by $\pi$ about two axes  ($\mathbb{Z}_2 \times \mathbb{Z}_2$). It is also well known that the ground state of the 1d spin-1 antiferromagnetic Heisenberg model is in the Haldane phase, and the SPT order actually survives for finite values of a possible (symmetry-preserving) uniaxial anisotropy term. For large values of the uniaxial anisotropy, though, the system undergoes a phase transition into a polarized trivial phase, which corresponds to a different projective representation of the Ê(in part on-site)\footnote{As such, bond-centered inversion does not look like an on-site symmetry since it implies the whole system. However, at the level of infinite-size MPS -- as we do here --, this can be accounted for locally by a transposition of the MPS bond indices, together with a possible permutation of the tensors within the MPS unit cell. In this sense,  at the level of practical implementation with MPS, this is also an `on-site symmetry".} symmetry group that protects the SPT phase. Such change in the projective representation is necessarily  discrete, and therefore it is only possible through a phase transition if the protecting symmetry is always preserved. However, if such symmetry is explicitly broken (e.g., by adding a staggered magnetic field), then the Haldane SPT phase can be adiabatically connected without closing the gap to other trivial product states. 

In this context, the existence of the so-called intermediate-Haldane phases was conjectured by Oshikawa more than 20 years ago already  \cite{iSPT}. In a broad sense, intermediate-SPT phases are effective topological spin-1 phases protected by symmetries that emerge in certain regions of the phase diagrams of higher-spin   quantum chains. The existence of such phases remained elusive for many years for numerical simulations, and could only be elucidated recently \cite{noyesiSPT, ouriSPT}. In fact, it has also been shown that all spin-1 SPT phases in 1d protected by $(\mathbb{Z}_2 \times \mathbb{Z}_2)+T $ symmetry can be realized as intermediate phases of a simple spin-2 quantum chain \cite{alliSPT}. 

Complementary, trivial phases of matter (i.e., those corresponding to a product state fixed-point under renormalization group) can also be protected by symmetries \cite{pollSPt}. Interestingly, this means that there are distinct trivial phases that cannot be adiabatically connected unless the symmetry protecting them is explicitly broken in the Hamiltonian. Or in other words, there are distinct product states that can be infinitely-close in limiting regions of a given phase diagram, yet separated by a quantum \emph{critical} point with diverging correlation length between them. This somehow counterintuitive statement was discussed in Ref.\cite{pollSPt}, where an explicit example of a spin-1 quantum chain with such a property was built. We call such phases \emph{symmetry-protected trivial} phases (SPt) \footnote{A clarification is in order here. In our convention, (i) SPT means that the corresponding quantum states cannot be fully disentangled without breaking the protecting symmetry (equivalently: they are not adiabatically connected to a product state by any path preserving the symmetry), and (ii) SPt means that the corresponding quantum states can be fully disentangled without breaking the protecting symmetry (equivalently: they are adiabatically connected to a product state by some path preserving the symmetry). In none of these two cases there is intrinsic topological order (or long-range entanglement) such as, e.g., in string-net models, so both cases are ``trivialÓ in this sense. This convention is slightly different to the one used by Wen\cite{wikiwen}, where SPT actually refers to both cases (i) and (ii) mentioned before. Our convention here is however compatible with the one in Ref.\cite{pollSPt}.}. Interestingly, it has also been shown that such phases can be adiabatically connected to non-trivial SPT phases such as the Haldane phase, yet the protecting symmetries in either case are different, as are their entanglement properties.  In particular, the SPt phases studied in Ref.\cite{pollSPt} were protected by a combination of site-centered inversion ($I_s$) and a $\pi$-rotation about the $z$ axis ($I' = I_s \times \mathbb{Z}_2$). 

Given all the above, a natural question arises: are there also intermediate-SPt phases in quantum spin chains? In this paper we give a positive answer to this question by analyzing an example of a spin-2 chain with a very rich phase diagram. More precisely, we show that for a given spin-2 quantum chain, effective intermediate spin-1 SPt phases emerge for some values of the parameters of the Hamiltonian, being also adiabatically connected to intermediate spin-1 SPT (non-trivial) phases. We characterize the phases of the model by studying quantities such as the entanglement entropy,  symmetry-related non-local order parameters, and 2-site fidelities. Our numerical analysis uses tensor network states \cite{tn1}, more precisely Matrix Product States (MPS) and the infinite Time-Evolving Block Decimation (iTEBD) algorithm, in order to approximate ground states of the system in the thermodynamic limit \cite{MPScan}. Moreover, we provide a field theory description in terms of two sine-Gordon bosons and one Majorana fermion for the Êpossible Ê quantum phase transitions between SPt phases. Combining this field-theory description with our numerical results, we conclude that the observed Êphase transitions are compatible with Conformal Field Theories (CFT) of central charge $c=1$, in perfect agreement with the results from Ref.\cite{pollSPt} (even though we cannot rule out the scenario of a first order transition for one of them).  Our findings show, for the first time, that  intermediate effective spin-1 SPt phases can also be realized in spin-2 quantum chains. 

This paper is organized as follows: in Sec.\ref{sec2} we review briefly the concept of SPt phase, and its connection to SPT phases. In Sec.\ref{sec3} the spin-2 quantum chain under study is introduced, and its properties are discussed. In Sec.\ref{sec4} we present our results. In particular, we discuss the field theory description and show numerical results for the entanglement entropy of half an infinite system, the entanglement entropy of a block at criticality, non-local order parameters, and 2-site fidelities. Finally, in Sec.\ref{sec5} we present our conclusions and outlook. 

\section{Trivial phases protected by point-group symmetries}
\label{sec2}

\subsection{Definition and properties}

By definition, a symmetry-protected trivial phase (SPt) is a phase for which all of its entanglement can be removed by local unitary transformations and, moreover, it is protected by certain point-group symmetries of the Hamiltonian. Even in the presence of the symmetry, the entanglement in the corresponding quantum state can be completely removed. Therefore, the phase is adiabatically connected to a trivial product state by a path that does not break the symmetry that is protecting it. This definition is to be contrasted with that of a symmetry-protected topological phase (SPT), where symmetry-preserving transformations cannot remove all the entanglement content in the quantum state or, equivalently, they cannot be adiabatically connected to a product state unless the symmetry is explicitly broken along the path. 

Importantly, it has been shown that in some 1d quantum spin chains such SPt phases can exist and are actually separated by quantum critical points between them \cite{pollSPt}. A bit counterintuitively, this means that one can be as close as one wants of two different product states, yet there is a quantum critical point between them with large (and even diverging!) amounts of entanglement. 

\subsection{Non-local order parameters from infinite MPS}
\begin{figure}
	\includegraphics[width=0.46\textwidth]{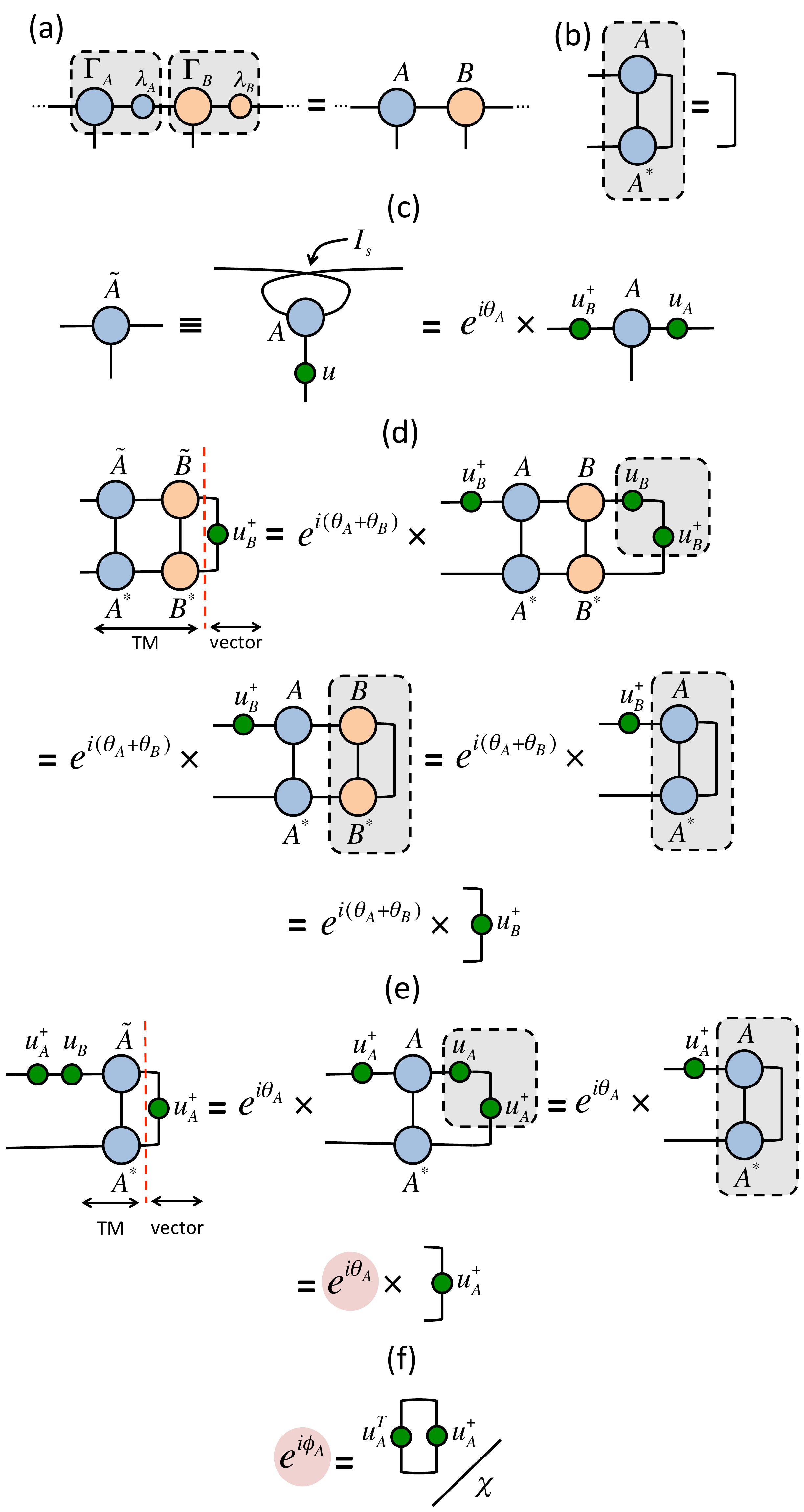}
	\caption{[Color conline] (a) MPS tensors $A$ and $B$ are defined as in the diagram; (b) tensor $A$ satisfies right-canonical condition (similarly for $B$); (c) action of the symmetry operator $I' = I_s \times \mathbb{Z}_2$ on tensor $A$ (similarly for $B$); (d) matrix $u_B^\dagger$ is the right eigenvector of a 2-site MPS transfer matrix (TM), with eigenvalue $\exp{i(\theta_A + \theta_B)}$ (similar conclusion by interchanging $A$ with $B$); (e) the phase $\exp{(i \theta_A)}$ (highlighted in red) can be extracted using $u_A^\dagger$ and a 1-site MPS transfer matrix built with tensor $A$ (similar conclusion by interchanging $A$ and $B$); (f) the phase $\exp{(i \phi_A)}$ (highlighted in red) can be extracted as shown in the diagram, where $\chi$ is the MPS bond dimension.}
\label{Fig1}
\end{figure}
SPt phases in 1d can be conveniently characterized by non-local order parameters which are easily computed in the MPS framework. Such calculation was first proposed in Ref.\cite{pollSPt}, and is briefly reviewed here for completeness for the case of an infinite-size MPS with translation invariance every two sites, which is  the relevant case for our purposes. The different steps are summarized in the diagrams of Fig.\ref{Fig1}. First, given an infinite MPS in canonical form \cite{MPScan} and with a two-site unit cell we can define two tensors $A$ and $B$ as in Fig.\ref{Fig1}(a) satisfying right-canonical conditions as in Fig.\ref{Fig1}(b). The symmetry operator $I' = I_s \times \mathbb{Z}_2$ acts on the MPS tensors as shown in Fig.\ref{Fig1}(c), where $u$ is a $\pi$-rotation generated by some hermitian operator,  which for our spin-2 case  will be defined at a later stage, and where the site-centered inversion $I_s$ simply takes the transpose of the bond indices in the MPS tensors $A$ and $B$. Matrices $u_A$ and $u_B$ correspond to representations of the symmetry operators acting on the MPS bond indices, and can be computed from the dominant right eigenvectors of 2-site MPS transfer matrices as shown in Fig.\ref{Fig1}(d). In turn, angles $\theta_A$ and $\theta_B$ can be extracted from the single-site MPS transfer matrices as shown in Fig.\ref{Fig1}(e). Finally, angles $\phi_A$ and $\phi_B$ are computed as in Fig.\ref{Fig1}(f). With all the data extracted in this way, one can define the SPt non-local order parameter following Ref.\cite{pollSPt} as
\beq
O^A \equiv e^{i (\theta_A - \phi_A)} = \pm 1 . 
\label{nonloc}
\eeq
Equivalently one can also define $O^B$ using $\theta_B$ and $\phi_B$. Let us stress at this point that the non-local order parameter depends on the definition of the $\pi$-rotation operator $u$. In our spin-2 case, we will use different operators in order to distinguish the different phases emerging in our model, as we shall see.

\section{The model}Ê
\label{sec3}

In this paper we consider the following spin-2 chain: 
\begin{equation}
H=\sum_{j}\left(\sum_{\gamma =1}^{4}J_{\gamma }(\vec{S}_{j}\cdot \vec{S}%
_{j+1})^{\gamma }+D(S_{j}^{z})^{2} -h_z(-1)^jS_j^z\right). 
\label{eq}
\end{equation}%
In the above Hamiltonian, $\vec{S}_j$ and $S^z_j$ are the usual spin-2 vector and Êits $z$-component , $D$ is a uniaxial anisotropy, $h_z$ is a staggered magnetic field, and $J_\gamma$ are values of bilinear, biquadratic, bicubic and biquartic couplings respectively for $\gamma = 1, \ldots, 4$. 
 
Some of the limiting regimes of this spin-2 quantum chain have been analyzed by us in previous papers\cite{Tu-2008, Scalapino-1998, ouriSPT, alliSPT}. Let us review them briefly in what follows. 

\subsection{$SO(5)$ point} 
As shown in Refs.\cite{Tu-2008, Scalapino-1998}, the case $J_{1}=-\frac{11}{6},J_{2}=-\frac{31}{180},J_{3}=\frac{11}{90},J_{4}=\frac{1}{60}$ with $h_z = D = 0$ is an exactly solvable point with $SO(5)$ symmetry and an exact MPS ground state \cite{Tu-2008,Scalapino-1998}. To see this, we notice that the model in this limit can be rewritten as 
\beq
H=2 \sum_{j}\left(P_{2}(j,j+1)+P_{4}(j,j+1) \right),
\eeq
where $P_{S_{T}}(j,j+1)$ projects onto total spin-$S_{T}$ states of neighboring sites $j$ and $j+1$. To identify the $SO(5)$ symmetry, we work in the
standard $S^{z}$ basis $|m\rangle $ ($m=\pm 2,\pm 1,0$)\ and define $SO(5)$
Cartan generators 
\beqa
L^{12}&=&|2\rangle \langle 2|-|-2\rangle \langle -2| \nonumber \\ 
L^{34}&=&|1\rangle \langle 1|-|-1\rangle \langle -1| .  
\label{ops}
\eeqa
By defining aditionally
\beqa
L^{15}&=&\frac{1}{\sqrt{2}}\left(|2\rangle \langle 0|+|0\rangle \langle -2|+\mathrm{h.c.}\right) \nonumber \\ 
L^{35}&=&\frac{1}{\sqrt{2}}\left(|1\rangle \langle 0|+|0\rangle \langle -1|+\mathrm{h.c.}\right), 
\eeqa
the SO(5) commutation relations 
\beq
[L^{ab},L^{cd}]=i\left(\delta _{ac}L^{bd}+\delta _{bd}L^{ac}-\delta _{ad}L^{bc}-\delta
_{bc}L^{ad}\right)
\eeq
fix the ten generators $L^{ab}$ $(1\leq a<b\leq 5)$. In the limit described previously, the Hamiltonian commutes with all ten operators $\sum_{j}L_{j}^{ab}$, and therefore has $SO(5)$ symmetry.

\subsection{$SO(5)$ point + $D$} 

As discussed in Ref.\cite{ouriSPT}, for $D>0$ the $SO(5)$ symmetry is explicitly broken down to $U(1) \times U(1)$. This is easy to see if we rewrite the uniaxial anisotropy term using 
\beq
(S^{z})^{2}=4(L^{12})^{2}+(L^{34})^{2}.
\eeq
This implies that operators $\sum_{j}L_{j}^{12}$ and $\sum_{j}L_{j}^{34}$ commute with the full Hamiltonian $H$ as well as with each other, implying a $U(1) \times  U(1)$ symmetry. Additionally, the Hamiltonian in this regime has several point-group symmetries including site-centered inversion $I_s$, time-reversal $T$, as well as a set of $\mathbb{Z}_{2}$ symmetries corresponding to $\pi$ rotations of the form 
\beq
u^{ab}=e^{i\pi L^{ab}}
\label{uab}
\eeq
for all $L^{ab}$. The $\mathbb{Z}_{2}$ operators
form a $(\mathbb{Z}_{2}\times \mathbb{Z}_{2})^{2}$ group, whose elements can be chosen as $\{\mathbb{I},u^{12}\}\times \{\mathbb{I} 
,u^{15}\} \times \{\mathbb{I} ,u^{34}\}\times \{\mathbb{I},u^{35}\}$. 

As shown in Ref.\cite{ouriSPT}, the model in this regime has two SPT phases: (i) one for low-$D$, called $SO(5)$-phase, with four-fold degenerate entanglement spectrum, and (ii) one for intermediate-$D$, corresponding to an intermediate spin-1 Haldane phase. The symmetries protecting these phases, as well as their different characterizations, are discussed in Ref.\cite{ouriSPT}. 

\subsection{Arbitrary $J_\gamma$ + $D$}Ê

Considering the effect of changing some of the values of $J_\gamma$ with respect from the $SO(5)$ point together with an uniaxial anisotropy, it was shown in Ref.\cite{alliSPT} Êthat one can actually find a host of intermediate spin-1 effective SPT phases. In particular, and as discussed in Ref.\cite{alliSPT}, in this case the Hamiltonian has a $(U(1) \times \mathbb{Z}_2) + T$ symmetry. The reduced discrete symmetry $(\mathbb{Z}_2 \times \mathbb{Z}_2) + T$ is known to protect up to 16 different possible SPT phases \cite{16p}. Four of these phases are typical of spin-1 chains (which includes the usual spin-1 Haldane phase). Quite remarkably, the four of them were found in the model for different values of $J_\gamma$ and $D$, with quantum critical points separating them corresponding to $c=2$ CFTs. 

\subsection{$SO(5)$ point + $D$ + $h_z$}Ê

This is the case considered in this paper, where both the uniaxial anisotropy $D$ as well as the staggered magnetic field $h_z$ can take finite values, but the $J_\gamma$ couplings are fixed to their $SO(5)$ point. Notably, the staggered field destroys all the symmetries protecting the SPT order, so we should expect these phases to break down. However, the model is still symmetric under site-centered inversion $I_s$ and $\pi$-rotations with either the $u^{12}$ or the $u^{34}$ operators from Eq.(\ref{uab}). As we shall see, we can define quantized SPt order parameters using the combined symmetries 
\beqa
I'_{12} &=&  I_s \times u^{12} \nonumber \\Ê
I'_{34} &=& I_s \times u^{34}, Ê
\label{sym}
\eeqa
which of course are also exact symmetries of the Hamiltonian in this regime. As we shall show here, the simultaneous presence of both protects \emph{three} trivial phases in our model, separated by quantum critical points corresponding to $c=1$ CFTs. One of these phases is adiabatically connected to the intermediate-Haldane phase for $h_z = 0$, and corresponds to an intermediate-SPt phase. 

\section{Results}Ê
\label{sec4}

Our aim here is to study the phase diagram of the spin-2 quantum Hamiltonian in Eq.(\ref{eq}), where the couplings $J_\gamma$ are fixed to the $SO(5)$ point, and both $D$ and $h_z$ can take finite values. For the sake of simplicity, we restrict ourselves to the case $D>0$ and $h_z > 0$. In our numerical simulations, we approximated the ground state of the system by an infinite MPS with a 2-site unit cell using the iTEBD algorithm. We used MPS of bond dimension up to $\chi = 100$, which turned out to be already sufficient for our purposes. In order to get a qualitative picture of the expected numerical results, we first discuss a field theory effective description of the Hamiltonian. As we shall see, such a description is in agreement with the numerical MPS results. 

\subsection{Field-theory description}

The model in Eq.(\ref{eq}) admits a simple field-theory description in terms of two bosonic sine-Gordon fields and one massive Majorana fermion field. This description is a generalization of the one in Ref.\cite{pollSPt}, and equivalent under refermionization, for Luttinger parameters $K_1 = K_2 = 1$, to the five-Majorana field theory introduced in Ref.\cite{ouriSPT}. The Hamiltonian density reads 
\beq
\mathcal{H}_{\mathrm{eff}} = \mathcal{H}_{b_1} + \mathcal{H}_{b_2} + \mathcal{H}_{f}, 
\label{eq:FT}
\eeq
with 
\beq
\mathcal{H}_{b_i} = \frac{v_i}{2 \pi}Ê\left( K_i (\partial_x \theta_i)^2 + \frac{1}{K_i} (\partial_x \phi_i)^2 \right) + g_i \cos{(2 \phi_i)} 
\eeq
the two bosonic sectors for $i = 1, 2$, and 
\beqa
\mathcal{H}_{f} = -iv \left(\xi _{R}\partial _{x}\xi
_{R}-\xi _{L}\partial _{x}\xi _{L} \right) - im\xi _{R}\xi_{L}
\end{eqnarray}
the Majorana fermionic sector. In these equations, $\theta_i$ and $\phi_i$ are dual fields satisfying $[ \phi_i(x), \theta_i(x') ] = i (\pi/2) ({\rm sgn}(x-x') + 1)$, and  $\xi_{R/L}(x)$ is a right/left-moving Majorana fermion field. The strategy to derive Eq. (\ref{eq:FT}) is essentially the same as in Refs.\cite{ouriSPT, pollSPt}, so we refer the reader to those references for further details.

The description of the quantum phase transitions is thus quite easy thanks to the field theory above. In particular, the Majorana field remains always gapped and therefore plays no role in the description of the quantum phase transitions. However, we include it so that the connection to the five-Majorana field theory in Ref.\cite{ouriSPT} is more transparent. We thus expect critical behaviors corresponding to the two massless bosons with $K_1, K_2  < 2$, and (i) $g_1 = 0$, and (ii) $g_2 = 0$. Both regions correspond, independently, to two different quantum critical lines with the central charge $c=1$ of a free massless boson. As we shall see, this is compatible with our numerical observations. 

\subsection{Entanglement entropy, spin-1/2 limit, and quantum criticality}

\begin{figure}
	\includegraphics[width=0.48\textwidth]{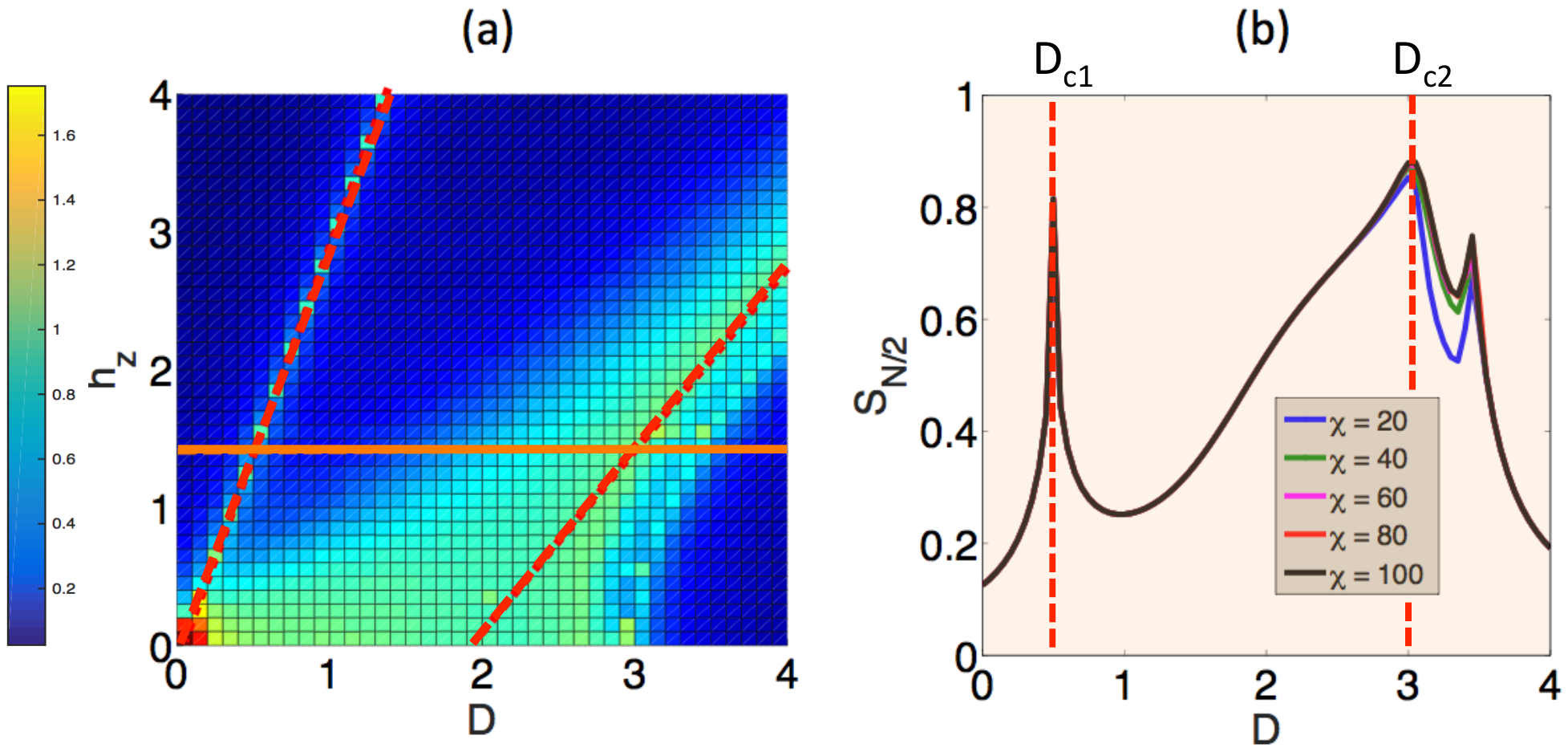}
	\caption{(a) Entanglement entropy of half an infinite chain for $\chi = 100$, aerial view. The two red dashed-lines correspond to the asymptotic laws in Eq.(\ref{as}) for the entropy maxima at $D \gg 1$. The horizontal orange line corresponds to $h_z = 1.5$, for which the entanglement entropy is plotted on the right panel. (b) Entanglement entropy of half an infinite chain for $h_z = 1.5$ and different bond dimensions $\chi$. Results largely overlap within our precision away from criticality. The two maxima on the right hand side tend to get closer as $\chi$ and $D$ increase (not shown).}
	\label{Fig2}
\end{figure}

As a first numerical calculation we computed the entanglement entropy of half an infinite chain, which can be extracted easily from the MPS tensors. In Fig.(\ref{Fig2})(a) we show our results (aerial view) in the $\langle D, h_z \rangle $ plane. The diagram clearly shows two regions of very large entropy which increases with the bond dimension. In the limit $\chi \rightarrow \infty$, we expect such entropies to diverge, thus pinpointing two quantum critical lines separating three phases. As we shall see, this interpretation is also consistent with the analysis of non-local order parameters. We also find that the maxima for the entanglement entropy in the critical lines follows the laws
\beqa
h_z &\sim& a_l + 3D  ~~~\rm (left)  , \nonumber \\ 
h_z &\sim& a_r + D  ~~~\rm (right)  , 
\label{as}
\eeqa
when $D \gg 1$, and for some constants $a_l$ (left) and $a_r$ (right). Such a behaviour is somehow expected\cite{strongcoupling}: for $h_z = D$, we find that the non-interacting piece of the Hamiltonian in Eq.(\ref{eq}) for an even site $j$ can be written using the usual spin-2 $z$-basis as 
\beq
D \Big( 2 (\ket{2}\bra{2} + \ket{-1}\bra{-1}) + 6 \ket{-2}\bra{-2} \Big).  
\eeq
with a similar result for odd sites but interchanging $2$ with $-2$ and $1$ with $-1$. This means that, for large enough $D$, the dominant low-energy physics will mostly happen in the 2-dimensional local subspace spanned by $\{ \ket{0}, \ket{1} \}$ for even sites, and $\{ \ket{0}, \ket{-1} \}$ for odd sites. So, locally, the low-energy physics in this limit resembles that of a spin-1/2 system. After projecting the interacting part of the Hamiltonian into this local basis, one gets a behavior similar to that of the spin-1/2 XXZ model, which in its critical region has $c=1$. Similarly, for $h_z = 3D$ we find that the non-interacting piece of Eq.(\ref{eq}) for an even site $j$ is given by  
\beq
D \Big( -2 (\ket{-2}\bra{-2} + \ket{-1}\bra{-1}) + 4 \ket{1}\bra{1} + 10 \ket{2}\bra{2} \Big), 
\eeq
again with a similar result for odd sites but interchanging $2$ with $-2$ and $1$ with $-1$. This could also suggest that for large $D$ the dominant  physics happens in a 2-dimensional local subspace, namely, the one spanned by $\{ \ket{-2}, \ket{-1} \}$ for even sites, and $\{ \ket{2}, \ket{1} \}$ for odd sites. Following the same argumentation as before, the relevant low-energy properties in this limit would then be that of some spin-1/2 XXZ model with $c=1$. Of course, this is just an intuitive argument and needs to be taken with care, but it would be in agreement with the field theory description in the case of having critical lines. Nevertheless, as we shall discuss, it is also plausible that one has a scenario with first order transitions, and therefore these limiting cases should be taken with care.  

\begin{figure}
	\includegraphics[width=0.4\textwidth]{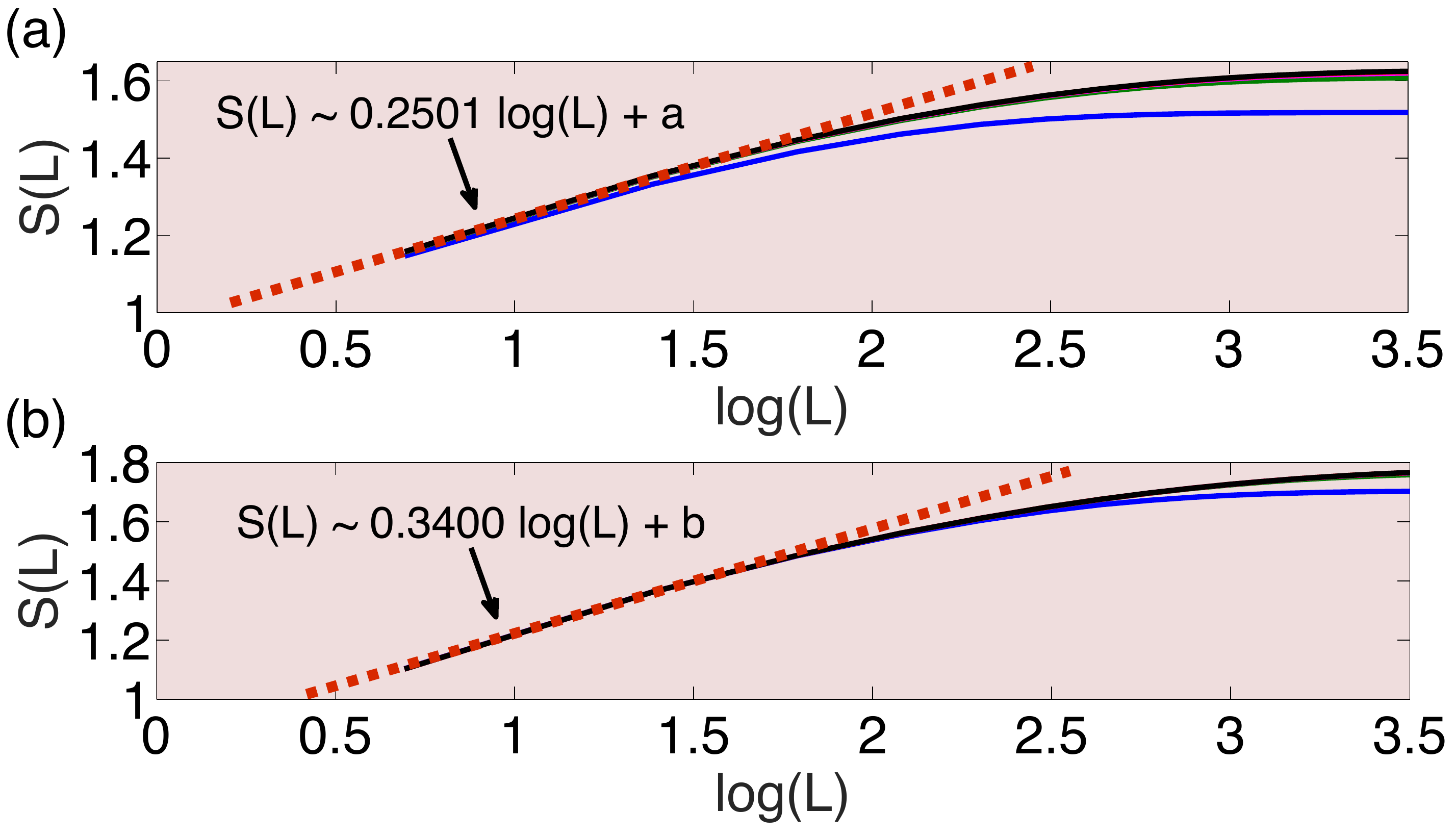}
	\caption{Entanglement entropy of a block of size $L$, at the two maxima of Fig.(\ref{Fig2}) for $h_z = 1.5$: (a) at $D_{c_1} = 0.5$; (b) at $D_{c_2} = 3$. The orange dashed line is a fit to the law in Eq.(\ref{cft}) in the scaling region. The values of the bond dimension $\chi$ are the same as in Fig.(\ref{Fig2}).}  
	\label{Fig3}
\end{figure}

Indeed, these results are also compatible with the numerical study of the scaling of the entanglement entropy. As shown in Fig.(\ref{Fig3}), the scaling of the entanglement entropy $S(L)$ of a block of size $L$ is compatible in both cases with the CFT scaling in the thermodynamic limit \cite{blockentropy}
\beq
S(L) \sim \frac{c}{3} \log L + c'_1
\label{cft}
\eeq
with central charge $c \sim 1$, once again in agreement with the previous considerations (in the above equation, $c'_1$ is a non-universal constant for an infinite chain, but for open boundary conditions it may also include the Affleck-Ludwig boundary entropy\cite{AL} --). More precisely, at $h_z = 1.5$ and for the first critical point $D_{c_1}$ we get $c = 0.2501 \times 3 \approx 0.75$, whereas for the second critical point $D_{c_2}$ we get $c = 0.34 \times 3 \approx 1.02$. The lower numerical central charge at $D_{c_1}$ can be explained by the fact that this is a fine-tuned quantum phase transition between two \emph{trivial} phases, corresponding to product states under RG. It is thus expected that the central charge is slightly underestimated by this fact, since trivial phases have very little entanglement. The same would be expected for the numerical central charge at $D_{c_2}$. However, this point is in the middle of a region with more entanglement (see Fig.(\ref{Fig2})), which apparently ``softens" the underestimation of $c$. Our conclusion is that both lines are compatible with a critical behavior, and seem to correspond to CFTs with central charge $c=1$. In any case, we would like to stress that within our accuracy we cannot rule out the possibility that the first phase transition is indeed first order, with a saturation of the entanglement entropy in the bond dimension of the MPS. This scenario would also be a plausible explanation of the observed low value for the central charge in our simulations.

\subsection{2-site fidelities}Ê
\begin{figure}
	\includegraphics[width=0.45\textwidth]{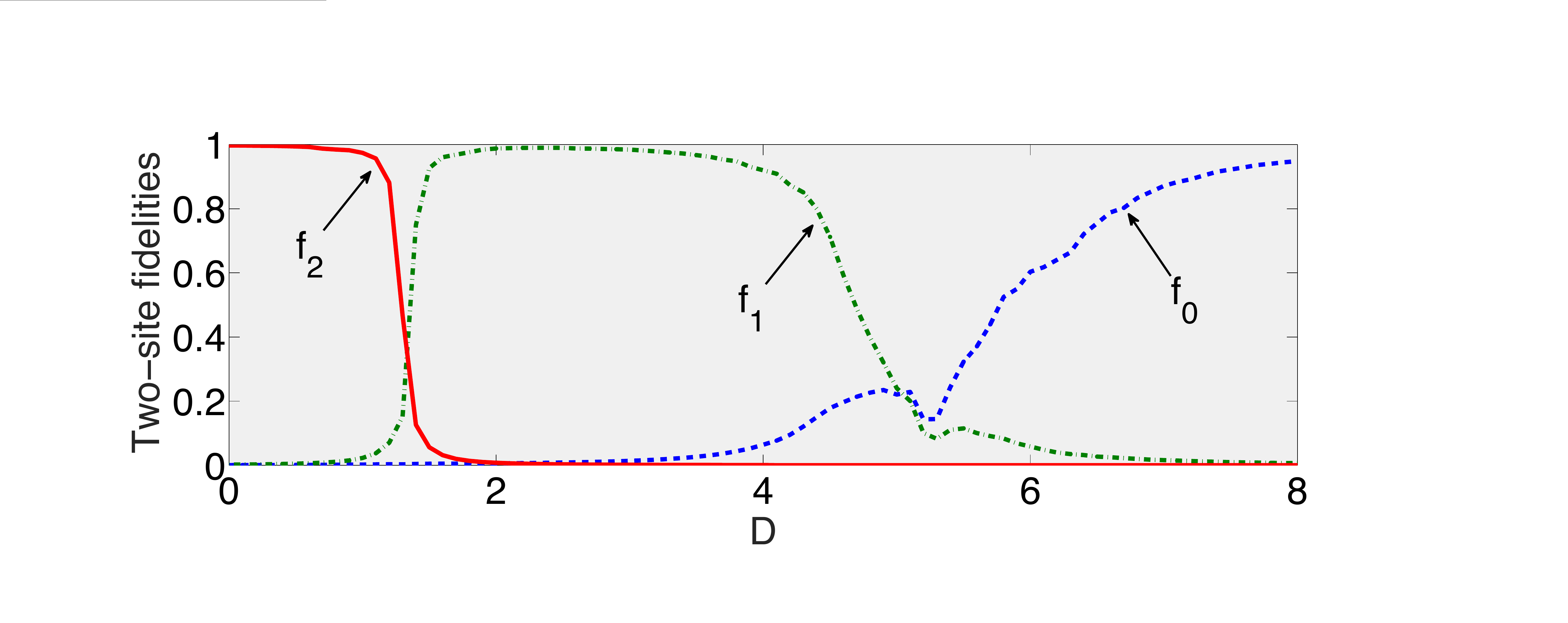}
	\caption{ Two-site fidelities $f_0, f_1$ and $f_2$ as in Eq.(\ref{fi}) at $h_z = 4$, clearly signalling 3 different trivial phases, for $\chi = 100$.}
	\label{Fig4}
\end{figure}

It is clear that introducing the staggered field term breaks explicitly the symmetries protecting the SPT phases at $h_z = 0$. Thus, for large values of $h_z$ one may expect distinct polarized phases. In order to identify them, we have computed the two-site fidelities
\beqa
f_0 &=& \frac{1}{2} \left(\bra{0} \rho_{odd} \ket{0} + \bra{0} \rho_{even} \ket{0}\right) \nonumber \\  
f_1 &=& \frac{1}{2} \left(\bra{1} \rho_{odd} \ket{1} + \bra{-1} \rho_{even} \ket{-1}\right) \nonumber \\  
f_2 &=& \frac{1}{2} \left(\bra{2} \rho_{odd} \ket{2} + \bra{-2} \rho_{even} \ket{-2}\right), 
\label{fi}
\eeqa
with $\rho_{odd}$ the 1-site reduced density matrix of the system for an odd site, and analogously for $\rho_{even}$. The results are shown in Fig.(\ref{Fig4}), and clearly show that the ground state of the system tends to be in three different trivial phases, which have the following product states as representatives for large $h_z$: $\ket{2,-2,2,-2, \ldots}$ for small $D$, $\ket{1,-1,1,-1, \ldots}$ for intermediate $D$, and $\ket{0,0,0,0, \ldots}$ for large $D$. 

\subsection{Protecting symmetries} 

\begin{figure}
	\includegraphics[width=0.46\textwidth]{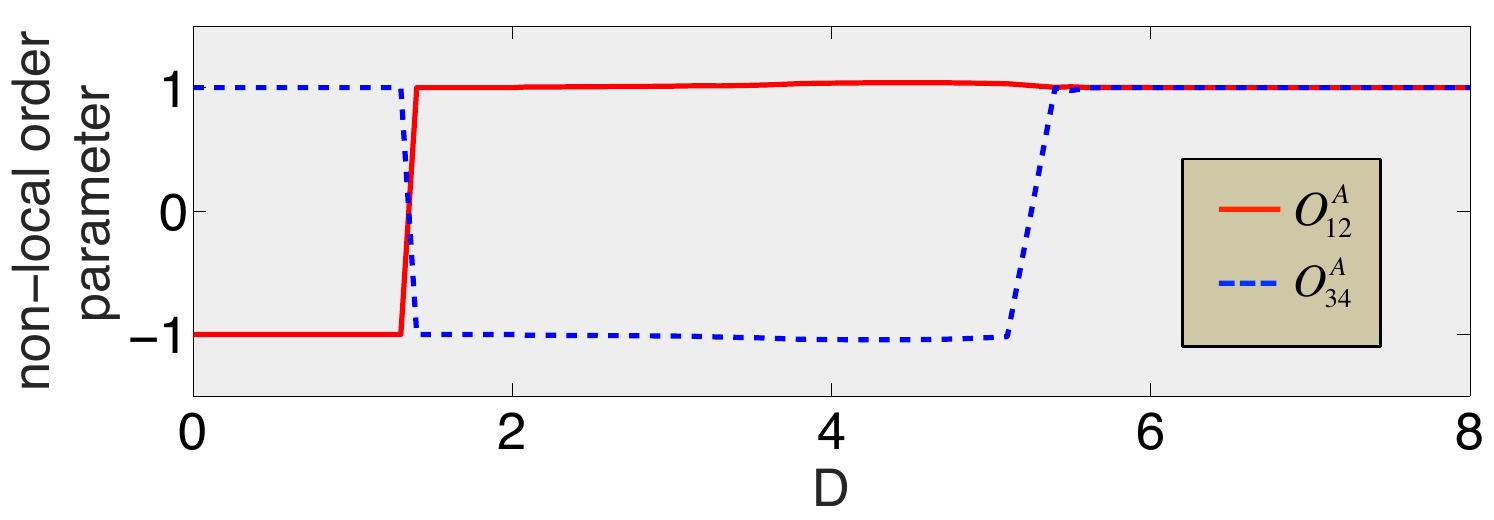}
	\caption{Non-local order parameters $O^A_{12}$ and $O^A_{34}$ as in Eq.(\ref{nonloc}) at $h_z = 4$, determining 3 different SPt phases, for $\chi = 100$.}
	\label{Fig5}
\end{figure}

Our numerical calculations also indicate that the three distinct trivial phases discussed above are in fact protected by symmetries. More specifically, the symmetries protecting the phases are those specified in Eq.(\ref{sym}), $I'_{12}$ and $I'_{34}$, consisting of site-centered inversion combined with a $\pi$-rotation according to operator $L_{12}$ or $L_{34}$ in Eq.(\ref{ops}). Defining two non-local order parameters $O^A_{12}$ and $O^A_{34}$ as in Eq.(\ref{nonloc}), we find that these have different values depending on the phase which allows us to distinguish the three phases, see Fig.(\ref{Fig5}) and Table \ref{tab}. 

\begin{table}[h]
\begin{center}
\begin{tabular}{||c|c|c|c||}
	\hline
	~~$O^A_{12}$~~ & ~~$O^A_{34}$~~ & ~~~Representative~~ & ~~~~~~~~Phase~~~~~~~~ \\
	\hline Ê
	 $-1$ & $+1$ & $\ket{2,-2,2,-2, \ldots}$ & small-$D$ \\Ê
	 $+1$ & $-1$ & $\ket{1,-1,1,-1, \ldots}$ & intermediate-$D$ \\Ê
	 $+1$ & $+1$ & $\ket{0,0,0,0, \ldots}$ & large-$D$ \\
\hline 
\end{tabular} 
\end{center}
	\caption{Different SPt phases found in our spin-2 quantum chain. The one for intermediate-$D$ corresponds to an intermediate-SPt phase.}
	\label{tab}
\end{table}

\section{Conclusions and outlook} 
\label{sec5}

Here we have shown for the first time the existence of an intermediate-SPt phase in a spin-2 quantum chain. We have characterized the phase diagram of our model by studying the entanglement entropy and its scaling, as well as non-local order parameters (pinpointing the protecting symmetries) and 2-site fidelities (pinpointing the relevant product states). Our numerical analysis, based on MPS in the thermodynamic limit, shows that Êthere may be critical regions  in the phase diagram compatible with $c=1$ CFTs, which we have characterized in terms of a theory with two sine-Gordon bosons and one Majorana fermion. In the large-parameter limit, such critical regions may be effectively described by some spin-1/2 XXZ models, as we have also shown. Concerning perspectives, it would be interesting to determine if a SPt phase with negative value of both non-local order parameters $O_{12}^A$ and $O_{34}^B$ could also exist in a simple quantum spin chain. Such a trivial phase may not be realizable if there are superselection rules forbiding it, and therefore may be SPT instead of SPt. Another interesting open problem is the characterization of SPt phases in 2d. We leave the study of these problems for the future. 

\acknowledgements

We acknowledge discussions with M. Oshikawa, F. Pollmann, and M. Rizzi, as well as with X.-G. Wen for pointing out a clarification concerning the definition of SPT phases. A. K. and R. O. acknowledge funding from the JGU and the DFG. H. H. T. acknowledges funding from the EU project SIQS.

\end{document}